\documentclass [amssymb, twocolumn, amsmath, showpacs]{revtex4-1}
\usepackage{graphicx,epsfig,amsfonts,amssymb}
\usepackage{bm}
\usepackage{times}
\begin{document}

\title{Fluctuation Relation for Heat Engines}

\author {N.~A. {Sinitsyn}$^{1,2}$}
\affiliation{$^1$Theoretical Division, Los Alamos National Laboratory, B258, Los Alamos, NM
87545} \affiliation{$^2$ New Mexico Consortium, Los Alamos, NM
87544, USA}

\date{\today}

\begin{abstract}
We derive the exact
equality, referred to as the fluctuation relation for heat engines
(FRHE),  that relates statistics of heat extracted from one of the two
heat baths and  the work per one cycle of a heat engine operation. Carnot's inequality of classical thermodynamics follows as a direct
consequence of the FRHE. 
 \end{abstract}


\date{\today} 

\maketitle

\begin{figure}[!htb]
\scalebox{0.25}[0.25]{\includegraphics{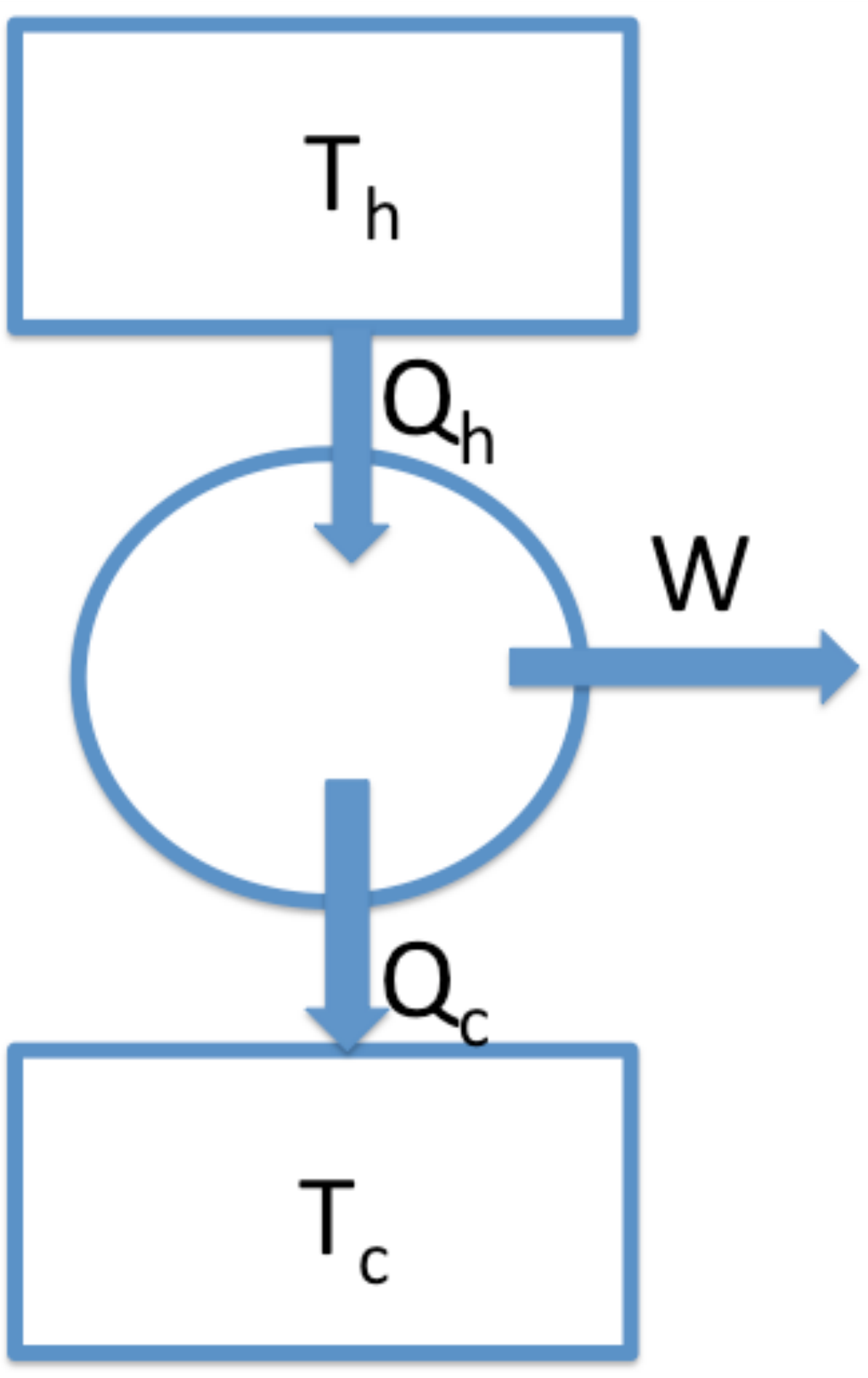}} \vspace{-3mm}
\hspace{1mm}   \caption{ Heat engine. Hot and cool baths (boxes) at
  temperatures, respectively, $T_{\rm h}$ and $T_{\rm c}$, are
  coupled to the central system (circle). $Q_{\rm h}$ is the heat
   that is absorbed from the hot bath  and $Q_{\rm c}$ is the heat that
   is released to the cool
  bath during the engine operation. Periodic changes of parameters of
  the central system coupled to baths and external fields lead to
  production of work, $W$, at a cost of heat
 fluxes from the hot to the cool baths.} 
\label{engine}
\end{figure}

The central thermodynamic result
for the efficiency of heat engines, which was established by Carnot in the 19th
century, reads
\begin{equation}
\frac{W}{Q_{\rm h}} \le 1-\frac{T_{\rm c}}{T_{\rm h}},
\label{carnot1}
\end{equation}
where $W$ is the work performed by a heat engine, $Q_{\rm h}$ is the heat
that was absorbed from the hotter bath during the cyclic
operation, and $T_{\rm c}$ and $T_{\rm h}$ are the absolute temperatures, respectively, of the
cooler and hotter baths, which are assumed constant during the process. 
The motive power of a heat engine originates from the passage of heat from a
hotter to a cooler heat bath, as shown in Fig.~\ref{engine}. Part of this heat flux can be converted to
useful work by means of a cyclic operation of a central system that is coupled to
both of the baths. When work is extracted in this way, there is always a heat
that is lost in the cooler bath. Carnot's inequality sets a limit on the useful work that
can be extracted from the absorbed power.

The inequality (\ref{carnot1}) was originally derived to describe
operations of macroscopic machines.  In modern times, a lot of
attention has been devoted to nonequilibrium thermodynamics of systems so small that thermal
fluctuations beyond the
realm of the Gaussian approximation can be observed and characterized. Stochastic behavior of mesoscopic heat engines is a growing field of modern research \cite{arnaut} with applications to nanostructures,
including nanocoolers \cite{ren-10prl} and molecular motors  \cite{astumian-rev11}. 
Surprisingly,
the stochastic behavior of strongly driven small systems  was found to
satisfy a universal constraint, as articulated in the Fluctuation Theorems (FTs) \cite{bochkov-77,evans-93,
  jarzynski-97prl,crooks-98,FT,gallavotti}. 
  
  FTs transform classical thermodynamic inequalities into equalities for exponents of 
thermodynamic variables. Eq. (\ref{carnot1}) is one of the classical inequalities. Consequently, it can be derived starting from a variety of known fluctuation relations \cite{jarzynski-97prl, jarzynski-99jsp}
but, to the best of author's knowledge, there is no written FT that  relates {\it exactly the same} variables as in (\ref{carnot1}) in a single exact expression.  The lack of such an equality may leave an impression that different  formulations of the 2nd law of thermodynamics, one of which is based on (\ref{carnot1}), are not completely equivalent when they are considered from the point of view of fluctuation relations.
In this article, we resolve this problem by deriving a FT, which we will call the FRHE, that relates the same variables as appear in (\ref{carnot1}) but now in the form of an exact equality.

The FRHE states that for an arbitrary heat engine  the work per cycle of the driving protocol and the 
heat absorbed from the hotter bath  per cycle are related by
\begin{equation}
\langle e^ { - Q_{\rm h} (1/T_{\rm c} - 1/T_{\rm h})
    +W/T_{\rm c} } \rangle =1, 
    \label{sinitsyn1}
\end{equation}
where the averaging is over many realizations of the thermodynamic cycle. 
Applying Jensen's inequality, $ \exp(\langle x \rangle ) \le \langle
\exp(x) \rangle $, Eq. (\ref{sinitsyn1}) leads to $\langle W \rangle
/T_{\rm c} \le \langle Q_{\rm h} \rangle (1/T_{\rm c} - 1/T_{\rm h})$, which is equivalent to
(\ref{carnot1}). Thus the primary importance of the FRHE is that it
relates the same variables as in the Carnot's theorem, thus promoting
the latter to the exact equality. 

Here we should specify the initial and the final conditions more clearly. We will assume that the central system is in the
thermodynamic equilibrium with the cooler bath at the beginning of the
protocol. The central system switches on the coupling to the hotter bath and
performs arbitrary changes of parameters
until finally it returns all parameters
to their initial values, including switching off the coupling to the hotter
bath, which ends the thermodynamic cycle.   Eq. (\ref{sinitsyn1}) becomes exact only for such protocols.

There are two limiting cases for which the FRHE (\ref{sinitsyn1})
reduces to the already known FTs. First, imagine that the system is completely decoupled from the hot bath. Then $Q_{\rm h}=0$, and 
(\ref{sinitsyn1}) becomes equivalent to the Bochkov-Kuzovlyov-Jarzynski equality \cite{bochkov-77,jarzynski-97prl}, 
$\langle \exp(W/T) \rangle = 1$, for a periodically driven system
coupled to a single heat bath. On the other hand, consider a system that does not perform any work. This happens e.g. when we simply  couple the  central system to the
heat baths without connecting the central system to other external
fields. Then the central system works just as a heat conductor between
two baths and Eq.~(\ref{sinitsyn1}) becomes equivalent to the so-called Exchange Fluctuation Theorem \cite{jarzynski-04prl}. 

We will not discuss the general domain of applicability of the FRHE and restrict its derivation 
only to the domain of stochastic thermodynamics \cite{jarzynski-rev09,FT-stoch-rev,seifert-06epl,Gaspard07,Kurchan,qian-11pre,ST-1,broek-10pre,parondo-09prl,ST-2,udo},
 in which continuous Markov chains are used to model fluctuations of physical systems. We note, however, that there are closely related expressions, such as Eq. (9) in Ref. \cite{jarzynski-99jsp} that were derived starting from 
 Hamiltonian equations of motion. This  indicates that extensions of the domain of the FRHE applicability should be possible.
 
 In our derivation of the FRHE, we assume that the dynamics of the central system is
 described by stochastic motion among a finite number $N$ of discrete
 states at chemical potentials $E_i$, $i=1,\ldots, N$. We assume that
 these discrete states are separated by barriers and that each barrier is strongly coupled to one of the heat baths. 
Each time the central system makes a transition from a state $i$ into a state
$j$ via a barrier it exchanges energy with a corresponding bath, which
forces the kinetic rates of transitions to have the Arrhenius
form. This guarantees that, at thermodynamic equilibrium, the system
state probabilities are described by the Boltzmann-Gibbs
distribution. For a transition through a barrier, which is coupled to
the hot/cool  bath, such a kinetic rate, $k_{ij}^{\rm h/c}$, is given by 
 $k_{ij}^{\rm h/c}=g_{ij} e^{E_i/T_{\rm h/c}}$, where $g_{ij}=g_{ji}$
 are parameters that characterize barrier sizes \cite{jarzynski-08prl}.
Every time such a transition occures, heat is transferred
between the central system and a corresponding bath. The corresponding change of the
energy of the
central system, $E_j-E_i$, is attributed to absorption/release
of energy from/to a hotter/cooler heat bath, i.e., respectively, either $\delta Q_{\rm h}=
E_j-E_i$ or  $\delta Q_{\rm c}=E_i-E_j$.

We consider protocols such that parameters $g_{ij}$ and
$E_i$ are the same at the beginning and at the end of the parameter driving. 
The probability, $P_F(X),$ of a trajectory, $X$, and the probability,
$P_B(\widetilde{X})$, of  its time-reversed counterpart,
$\widetilde{X}$, in the same system driven by a time-reversed protocol, satisfy the relation \cite{crooks-98,jarzynski-rev09,udo}
\begin{equation}
\frac{P_F(X) e^{S(X)}}{P_B( \widetilde{X})} = \frac{f_0^{\rm
    eq}}{f_{\tau}^{\rm eq}}.
\label{traj}
\end{equation}
Here $S(X)=- {\rm ln}\prod_r (k_{ij}(t_r)/k_{ji}(t_r))$, where the
product is over all transition time moments, enumerated by index $r$, along a
trajectory $X$, and where $t_r$ is the time of the transition from the
state $i$ into the state
$j$ along $X$; $k_{ij}(t_r)$ is the kinetic rate of this transition
and  $k_{ji}(t_r)$ is the kinetic rate of the transition in the
opposite direction at time $t_r$.
$f_{0}^{\rm
    eq}$ and $f_{\tau}^{\rm
    eq}$ are the equilibrium probabilities of having, respectively,
  initial and final (at time $\tau$ of the end of the protocol)
  states of the trajectory when the central system is decoupled from
  the hot bath.  Reformulating (\ref{traj}) in terms of $E_i$ and
  $g_{ij}$, we find that
\begin{equation}
\frac{P_F (X) } {P_B( \widetilde{X})} e^{Q_{\rm h}/T_{\rm h} - Q_{\rm c}/T_{\rm c}} = \frac{f_0^{\rm
    eq}}{f_{\tau}^{\rm eq}}, 
\label{traj2}
\end{equation}
where $Q_{\rm h/c}=\sum_{r} \delta Q_{\rm h/c}(t_r)$
 are total heat amounts transferred from/to hot/cool baths to/from
 the central system, assuming that at the beginning of the cycle $Q_{\rm h/c}=0$.

The energy balance between the work, $W$, the heat fluxes, $Q_{\rm h}$,
$Q_{\rm c}$, and the change of the internal energy of the central system, $\Delta E= E(\tau)-E(0)$, is generally given by  
\begin{equation}
Q_{\rm h}-Q_{\rm c}=W+ \Delta E.
\label{cons1}
\end{equation}

Our goal is to relate $Q_{\rm h}$ and $W$.
One can think that since Eq.(\ref{cons1}) is the only exact constraint among 4 variables $Q_{\rm h}$, $Q_{\rm c}$, $W$ and $\Delta E$, further fluctuation relations should involve at least 3 of those variables.
However, when averages of exponents are considered, additional cancellations happen.  Eliminating $Q_{\rm c}$ from (\ref{cons1}) and substituting the result
in (\ref{traj2}) we find 
\begin{equation}
\frac{P_F (X) } {P_B( \widetilde{X})}
 e^{Q_{\rm h}(1/T_{\rm h} -  1/T_{\rm c}) +W/T_{\rm c}} = 1, 
\label{traj3}
\end{equation}
where we used the fact that due to the detailed balance conditions,
$e^{E(0)/T_c} f_0^{\rm eq} = f_{\tau}^{\rm eq} e^{E(\tau)/T_c}$.

 Let
$\rho (Q_{\rm h},W)$ be the probability density to observe given values of
$Q_{\rm h}$ and $W$. The standard manipulations
\cite{jarzynski-rev09} then give us
\begin{equation}
\begin{array}{l}
\rho (Q_{\rm h},W) = \int dX P_F(X) \delta (W-W^F(X)) \delta (Q_{\rm
  h} -Q_{\rm h}^F(X)) = \\
\\  
e^{-Q_{\rm h}(1/T_{\rm h} -  1/T_{\rm c}) 
  -W/T_{\rm c}} \int d\widetilde{X} P_B(\widetilde{X}) \delta
(W+W^B(\widetilde{X})) \delta (Q_{\rm h} + \\
\\
 Q_{\rm h}^B(\widetilde{X}))=
e^{-Q_{\rm h}(1/T_{\rm h} -  1/T_{\rm c})   -W/T_{\rm c}} \rho (-Q_{\rm h},-W),
\end{array}
\label{rho}
\end{equation}
where $Q_{\rm h}^F(X) =-Q_{\rm h}^B(\widetilde{X})$ are the
quantity of heat 
absorbed from the hotter bath during forward and backward protocols in
motion 
along, respectively, the forward and the backward trajectories. Analogously we
denote by $
W^F(X) =-W^B(\widetilde{X})$ the amount of
work done.
Summing in (\ref{rho}) over all possible $Q_{\rm h}$ and $ W$,
we obtain the FRHE (\ref{sinitsyn1}), which completes its derivation.

{\it In conclusion,} we derived a fluctuation relation that couples
the work performed by a heat engine with the heat flux from the hotter
heat bath. The equality describes performance of an arbitrary heat
engine that can be driven by arbitrarily fast and strong
time-dependent fields. We also allowed both heat baths to be coupled
strongly to the central system at the same time, except the beginning
and the end of the driving protocol. The FRHE directly
leads to  Carnot's inequality for macroscopic heat engines and, in this sense, it may be regarded as generalization of this crucial
thermodynamic principle. Certainly, at present,  Carnot's
inequality remains the true fundamental law of physics because the
full domain of applicability of the FRHE in the real world remains to be understood. 
FTs have already been used for practical purposes, including the
development of accelerated free energy sampling algorithms
\cite{ft-algorithms}. The FRHE should similarly help
the research on optimization and control of nanoscale device
structures such as electric coolers and molecular motors \cite{astumian-rev11}.

{\it Acknowledgment}. Author thanks Allan Adler and Maryna Anatska for useful discussions. The work at
LANL was carried out under the auspices of the National Nuclear
Security Administration of the U.S. Department of Energy at Los
Alamos National Laboratory under Contract No. DE-AC52-06NA25396. It
is also based upon work supported in part by the National Science
Foundation under ECCS-0925618 at NMC.


\end{document}